\documentclass[12pt]{iopart}
\usepackage{graphicx}
\usepackage{color}

\begin{document}

\title[`Nanobig' nanorods]{The `nanobig rods' class of gold nanorods: optimized dimensions for improved \textit{in vivo} therapeutic and imaging efficacy}

\author{Constantin Ungureanu$^{1,}$\footnote[1]{$c.ungureanu@tue.nl.$}, Gerben A. Koning,$^{2}$  Ton G. van Leeuwen$^{1,3}$ and Srirang Manohar$^{1}$\footnote[2]{$s.manohar@utwente.nl.$}}

\address{$^{1}$Biomedical Photonic and Imaging Group, MIRA Institute for Biomedical Technology and Technical Medicine, Faculty of Science and Technology, University of Twente, P.O. Box 217, 7500AE Enschede, The Netherlands}
\address{$^{2}$Laboratory Experimental Surgical Oncology, Section Surgical Oncology, Department of Surgery, Erasmus Medical Center, Rotterdam, The Netherlands}

\address{$^{3}$Biomedical Engineering and Physics, Academic Medical Center, University of Amsterdam, PO Box 22700, 1100 DE Amsterdam, The Netherlands}

\date{\today}

\begin{abstract}

Currently, gold nanorods can be synthesized in a wide range of sizes.
 However, for intended biological applications gold nanorods with approximate dimensions  50nm x 15nm are used.
We investigate by computer simulation the effect of particle dimensions on the optical and thermal properties in the context of the specific applications of photoacoustic imaging.
 In addition we
 discuss the influence of particle size in
overcoming the following biophysical barriers when administrated \emph{in vivo}: extravasation, avoidance of
uptake by organs of the reticuloendothelial system, penetration
through the interstitium, binding capability and uptake by the
target cells. Although more complex biological influences can be
introduced in future analysis, the present work illustrates
that larger gold nanorods, designated by us as ``nanobig rods", may perform
relatively better at meeting the requirements for successful
\emph{ in vivo} applications compared to their smaller counterparts which are
conventionally used.

\end{abstract}





\maketitle

\section{Introduction}

One of the most important property of gold nanoparticles (AuNP) is the intense absorption of light at specific wavelengths, due to the phenomenon of localized surface plasmon resonance (LSPR). Additionally, these particles are relatively biological inert and by using different conjugation strategies, such as PEGylation ~\cite{Lankveld2011,Jokerst2011,Green2011}, where antibodies are attached on their surface, they can be used to target specific biomolecules.

SPR occurs when light of a specific frequency sets free electrons of the AuNP surface into collective resonant oscillations (plasmons). The resonant frequency is uniquely defined by particle size, shape~\cite{Huang2009,Jain.ACR2008,Perez-Juste2005} and dielectric enviroment.
At resonance, the interaction of the incident light and the AuNP is high, leading to narrow absorption and/or scattering peaks in the spectra. In asymmetric AuNP, plasmons can be created along the different axes of the particle, giving rise to multiple plasmon bands in the spectra. For example in gold nanorods (AuNR) ~\cite{Prescott2006,Ungureanu.JAP2009}, resonant oscillations can occur along the short axis and the long axis, causing a transverse peak (TP) and a longitudinal peak (LP) respectively in the spectra. The TP is situated in the green region of the spectrum; the LP is red-shifted and tunable with aspect ratio~\cite{Huang2009} to occur in near infrared region (NIR). The NIR wavelength region is interesting for applications in tissue, since absorption ($\mu_{a}$) and scattering ($\mu_{s}$)
coefficients of tissue are relatively low in this region, allowing high
penetration of light for imaging into tissue (higher than 1 cm)~\cite{Li2009e,Qin2012,}.
 Since the LP-driven absorption peak of AuNRs occurs in the NIR, these particles can be used as imaging contrast agents especially in photoacoustic imaging
 ~\cite{Manohar2011,Huang2011,Su2011,Yeager2012,Mallidi2011,Tong2009}.

Under illumination conditions of thermal and stress confinement, the local temperature rise produces ultrasound waves by photoacoustic effect; measurement of the ultrasound transients allows detection and visualization of the disease site.
 The temperature rise around irradiated particles can also produce therapeutic effects. CW (continous wave) light irradiation can be used to cause cell death in the process of hyperthermia~\cite{Li2009,Emelianov.Nanolett2009,Huang2009, AmyL.Oldenburg2009}.

 For these biomedical applications, various methods have been researched in pre-clinical studies whereby the NPs can be functionalized
~\cite{Tiwari2011}  by conjugating them with antibodies, thereby imparting them with the capability to target disease sites such as cancer~\cite{Manohar.IJBI2007,Manohar2011}.

The ability to detect the disease or to affect a complete
therapeutic action is dependent on the extent of interaction
that the NPs will have with light, phenomenon for which NPs can be
tailored by appropriate choice of physical features such as size and
aspect ratio. Also important, is extent to which a therapeutically
relevant concentration accumulates homogeneously throughout the
disease area. This requires design of physical and biochemical
features of the NPs, which calls for some understanding of the physiology in normal and tumor tissue, and in
their respective vasculature.

In general, for biomedical application of NPs the following steps are
required for deployment of NPs for diagnostic or therapeutic purposes.
\begin{enumerate}
\item synthesis of the NPs
\item NP bioconjugation with disease specific antibodies (mAb)
\item topical or systemic administration of mAb-NPs
\item circulation in blood stream
\item extravasation at disease site through leaky vasculature
\item transport in tumor
\item binding to the targeted cells
 \item triggering of NPs present at the diseased site (tumor) with light,
for detection or therapy.
\end{enumerate}

Currently AuNRs can be synthesized with large variations in
physical dimensions (length, width and aspect ratio) each variant
having specific optical properties~\cite{Perez-Juste2005}.
Typically, the entire ``optical diagnostic and therapeutic
window''in the NIR spectrum can be covered by AuNRs with aspect
ratios (a.r) ranging from 2.5 to 5, and effective radii ($r_{eff}$) from
5 nm to 35 nm~\cite{Harris2008,Prescott2006,Perez-Juste2005}. The $r_{eff}$ of a AuNR is the radius of a sphere
having the same volume as the particle~\cite{Draine}.

The immediate question is: which particle among these are optimum for diagnostic and/or therapeutic purposes?
The goal of this article is to provide a possible solution to this problem.

Using computer simulations and analysis of
data reported in literature, we find that larger AuNRs
(larger $r_{eff}$), while still preserving appropriate a.r, can largely meet the requirements for successful use in biological applications.
These particles designated by us ``nanobig rods" have better physical, optical and thermal
properties compared with the commonly used gold nanorods which
have $r_{eff}$ smaller than half of mean free path of electrons in gold. Next to the improved optical responses, we  also consider the
effects of particle dimensions can have on \textit{in vivo}
behavior such as on extravasation, uptake by cells and thermal stability in laser field.

\section*{Materials and methods}
\subsection*{Simulation of optical properties of AuNRs}
The DDSCAT 6.1~\cite{Draine} package (an implementation of the Discrete
Dipole Approximation method)  was used to simulate the optical
properties of AuNRs. The method discretizes a particle into
dipoles, and the electromagnetic field scattered by the
nanoparticle is calculated taking into account dipole-dipole and
dipole-light interaction. The approach allows the interaction of
light with arbitrary shaped particles to be modeled and
simulated, with a knowledge of parameters such as the dielectric
function of the material, refractive index of media and particle
orientation relative to incident electromagnetic field. For
simulating AuNR, we used the dielectric function of bulk
gold~\cite{Palik1991}, and water with refractive index of 1.33 was
considered as embedding medium~\cite{Ungureanu.JAP2009}.

\section*{Criteria for choosing sizes of the gold nanorods}
\subsection*{Optical properties}
In \emph{in vivo} biomedical applications collections of particles are
 involved. The absorption coefficient $\mu_{a}$~\cite{Li2009e} of such an ensemble is wavelength dependent~\cite{Ungureanu2010} and is calculated as the product of particle concentration
(\emph{N}) and absorption cross-section ($C_{abs}$) of the particle ($\mu_{a}=N \times C_{abs})$ where:
\begin{equation}\label{tre}
          C_{abs}=\pi r_{eff}^{2} Q_{abs}
\end{equation}

with $Q_{abs}$ the absorption efficiency of the particle.
For maximizing light interactions with the NRs embedded in tissue,
the following requirements for the NR need to be fulfilled:
\begin{itemize}
    \item the LP peak has to be located in the NIR spectral region where background tissue optical properties, $\mu_{a}$ and
    $\mu_{s}$, are lower than in visible range~\cite{Vo-Dinh2002, Franceschini1999},
    \item the optical interaction coefficients, $\mu_{a}$ or $\mu_{s}$, should be as high as
    possible, to increase the light induced effect necessary for
    detection or therapy.
\end{itemize}

AuNRs are currently synthesized using various
modifications~\cite{Perez-Juste2005} to a seed-mediated
silver-assisted growth protocol~\cite{Nikoo.ChemMat2003} resulting
in variously sized rods. We modelled a wide range of AuNRs with
$r_{eff}$ (5-35 nm) and aspect ratios (AR) (2.5 - 4) to include
these different particles.

For \textit{in vivo} optical imaging and photothermal
applications, 800 nm is a wavelength that is typically used~\cite{Li2009e,E.Dickerson2008}. In
this NIR region, tissue has relatively low $\mu_{a}$ and $\mu_{s}$
Further, sources of 800 nm in the ns regime by pumping OPO
crystals are readily available for photoacoustic imaging.  CW laser sources with emission wavelength at 800 nm are also reported in studies about hyperthermic effect~\cite{Choi2011,Cole2009}.  For these
reasons we make comparisons between AuNRs which show their LP peaks
around this wavelength.

\begin{figure}[!h]
\centering
\includegraphics [width=60mm]{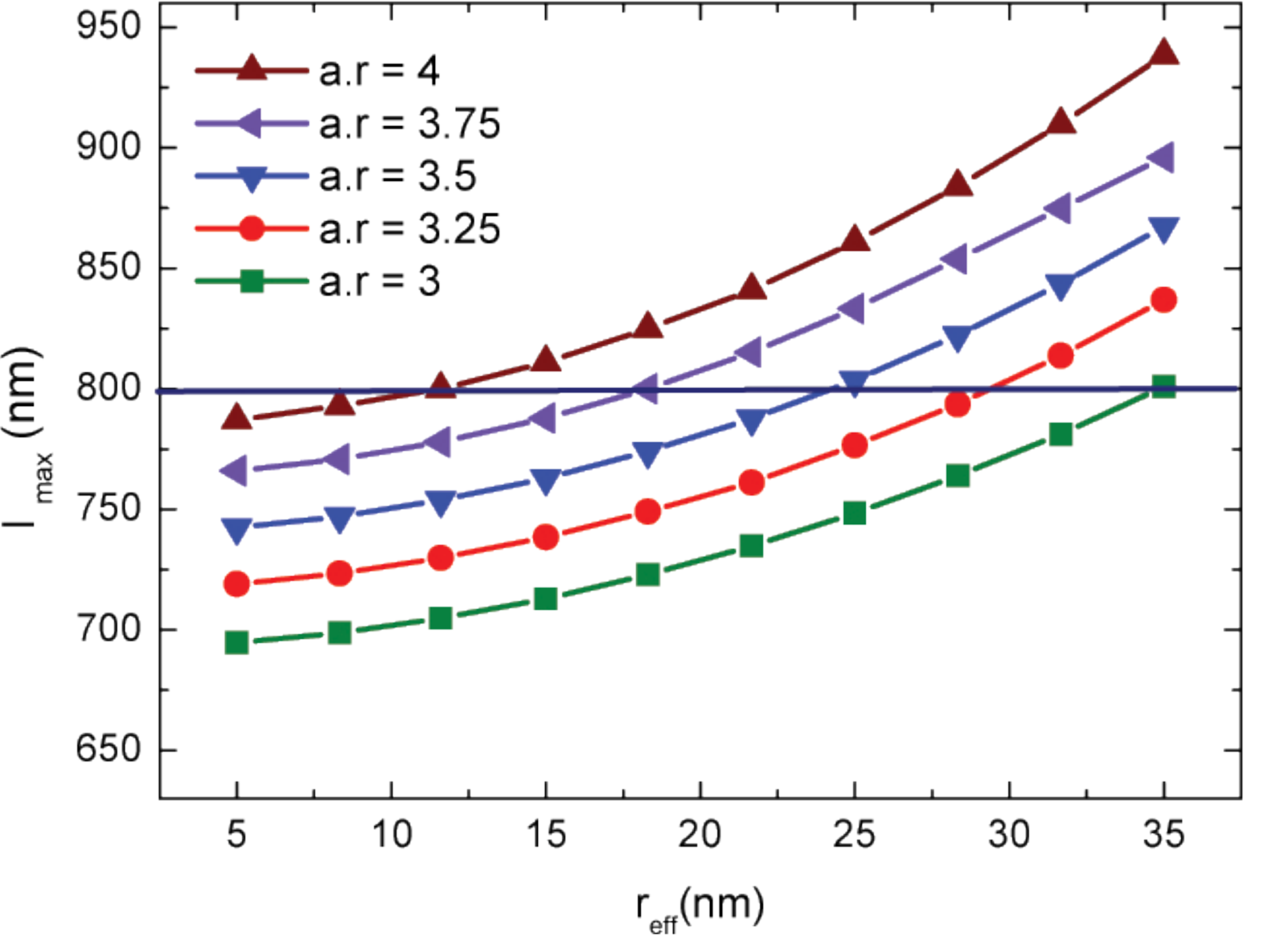}
\caption {Location of the longitudinal plasmon (LP) peaks
simulated for AuNRs with aspect ratios (AR) 3, 3.25, 3.5, 3.75 and
4, as a function of effective radius.}
 \label{fig:one}
\end{figure}
Figure~\ref{fig:one} shows as a function of $r_{eff}$ the spectral
locations of the LP absorption peaks for AuNRs with AR 3, 3.25,
3.5, 3.75 and 4. As observed earlier~\cite{Jain2006c},
not only the AR but also the volume of the NR determines the
position of the LP peak. The LP peaks red-shift with increasing
$r_{eff}$, but the region around 800 nm is covered only by AuNRs
with AR between 3 and 4, possessing $r_{eff}$ between 11
and 35 nm.

\begin{figure}[!h]
\begin{minipage}{.45\linewidth}
\centering
\includegraphics[width=\linewidth]{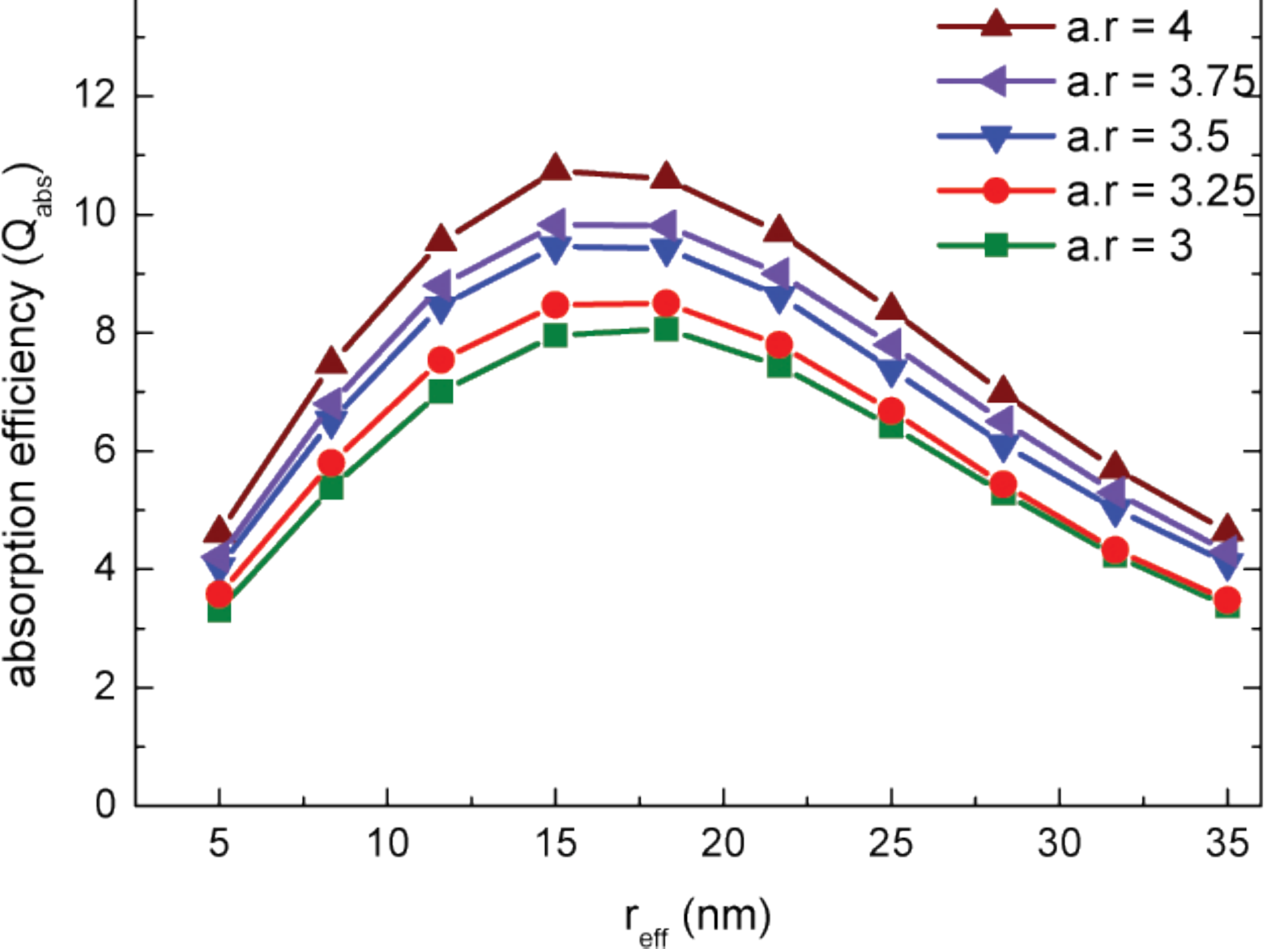}
\centering{(a)}
\end{minipage}
\begin{minipage}{.45\linewidth}
\centering
\includegraphics[width=\linewidth]{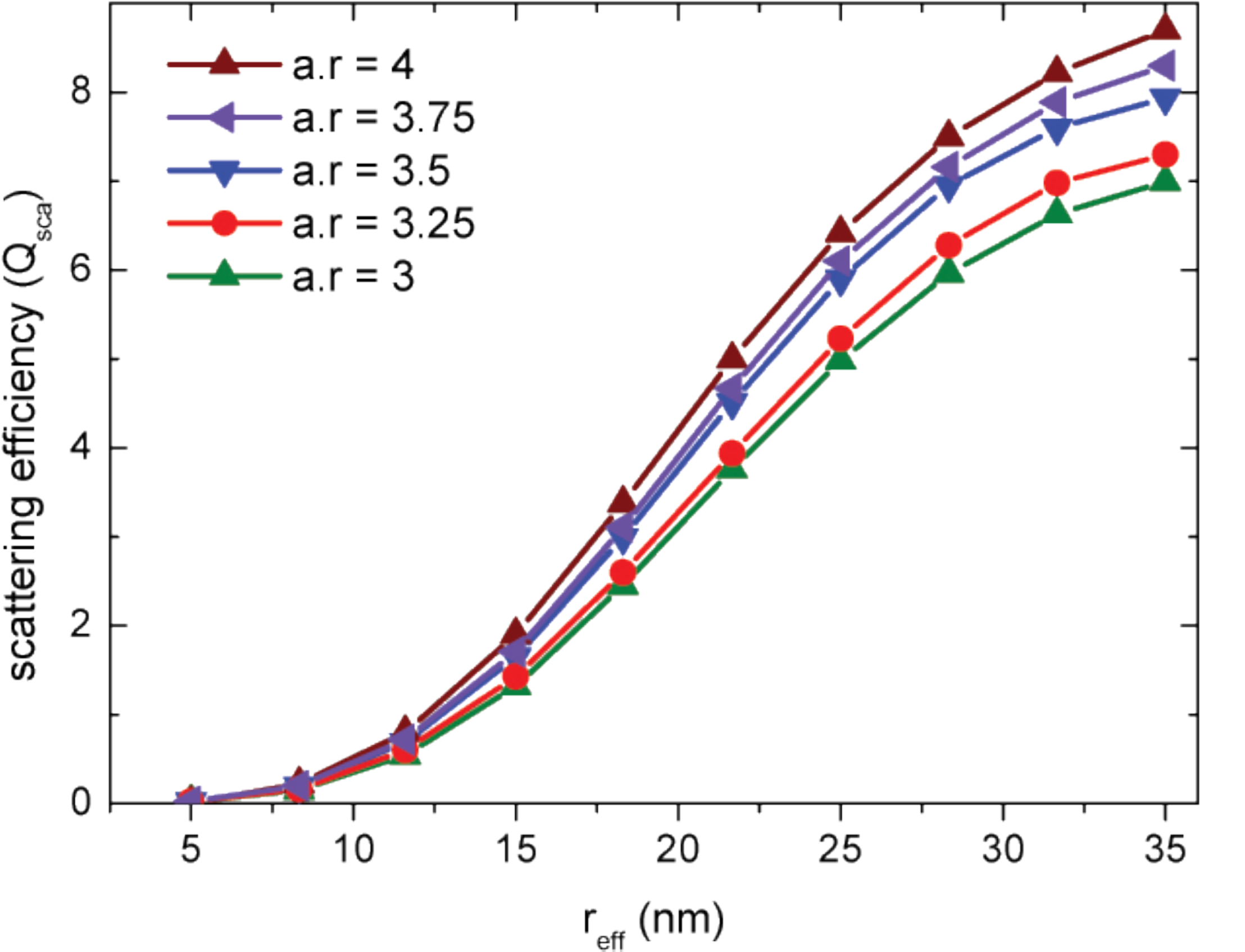}
\centering{(b)}
\end{minipage}
\caption {Simulated absorption efficiency  Q$_{abs}$ (a), and
scattering efficiency  $Q_{sca}$ (b), at the LP as a function of
effective radius for aspect ratios from 3.0 to 4. Q$_{abs}$ passes
though a maximum at about 17.5 nm. $Q_{sca}$ increases with
effective radius reaching saturation levels above 35 nm.}
  \label{fig:two}
\end{figure}

The values of simulated absorption ($Q_{abs}$) and scattering
($Q_{sca}$) efficiencies at the LP peaks for a.r 3 to 4 as a function
of $r_{eff}$ are shown in figure~\ref{fig:two} (a) and (b).
Irrespective of a.r, $Q_{abs}$ values have a peak around
$r_{eff}=17.5$ nm. The diameter of such particle is close to the mean free path of electrons in gold ($\approx$ 42nm~\cite{Kooij2006}).

 $Q_{sca}$
increases with increasing $r_{eff}$ but NRs with $r_{eff}<$ 10 nm
can be considered pure absorbers as $Q_{sca}$ is negligible in
comparison with $Q_{abs}$. For $r_{eff}>27$ nm (see
figure~\ref{fig:two}), scattering is larger than absorption.
 This is the consequence of radiation damping
effects which occur in larger particles~\cite{Link1999}, a
behavior also seen in gold nanospheres~\cite{Yin2006}.

\begin{figure}[!h]
\centering
\includegraphics[width=\linewidth]{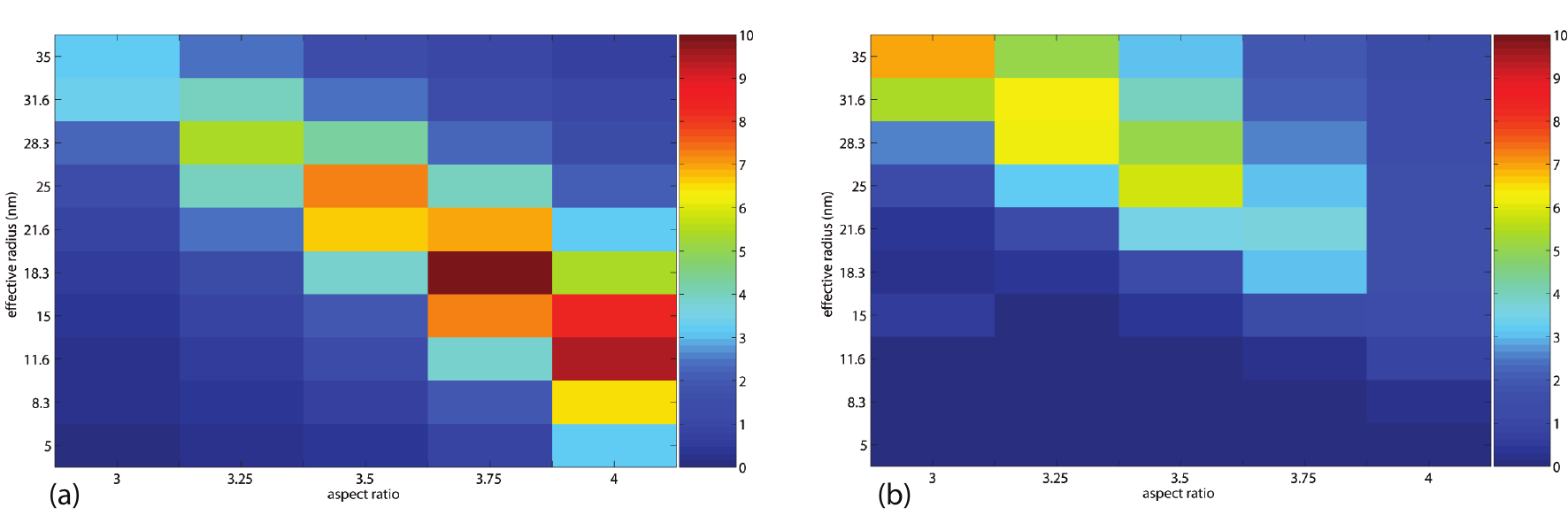}
\caption {Simulated absorption efficiency (a) $Q_{abs}$, and
scattering efficiency (b)  $Q_{sca}$, at 800 nm as functions of
effective radii ($r_{eff}$) and aspect ratios (a.r).}
  \label{fig:three}
\end{figure}

Figure~\ref{fig:three} depicts the optical properties in the form of  $Q_{abs}$ and  $Q_{sca}$ values at 800 nm for
particles with a.r between 3 and 4, and $r_{eff}$ between 11 and 35 nm. From
figure~\ref{fig:three}(a) we can identify an optimal combination
of a.r$=3.75$ and $r_{eff}=$17.5 nm for obtaining the LP peak at
800 nm with the highest $Q_{abs}$. As shown above, for thermal
response upon irradiation, $\mu_{a}$
is the most important parameter.

\begin{figure}[!h]
\centering
\includegraphics[width=\linewidth]{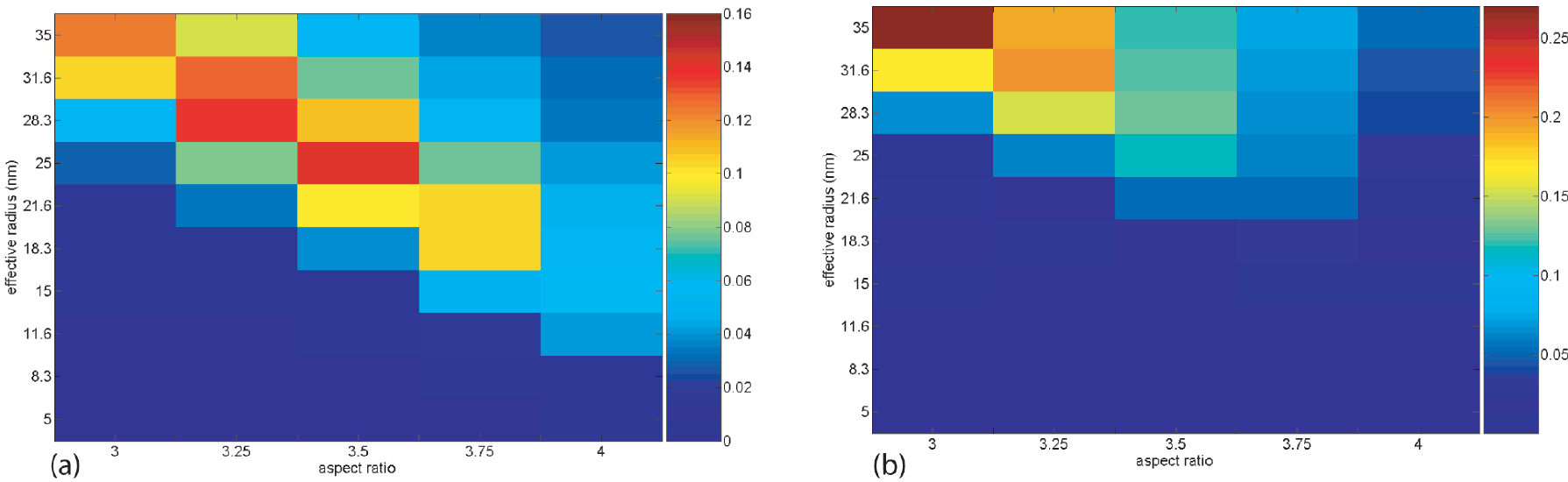}
\caption {Simulated absorption coefficient $\mu_{a}$ (cm$^{-1}$)
(a) and scattering coefficient $\mu_{s}$ (cm$^{-1}$) (b)
calculated at 800 nm for a collection of 10$^{9}$/ml rods as a
function of $r_{eff}$ and a.r at 800 nm.}
  \label{fig:four}
\end{figure}

We calculated further the $\mu_{a}$ for solutions containing particles simulated in figure~\ref{fig:one}, using the  $Q_{abs}$ and  $Q_{sca}$  from figure~\ref{fig:three} at the same particle concentration (10$^{9}$/ml).
The result displayed in figure~\ref{fig:four} shows that collections of particles with the combination of a.r of 3.75 and $r_{eff}=$17.5 nm are not actually optimal, with particles with a.r of 3.5 and $r_{eff}=$25 nm possessing higher $\mu_{a}$  and $\mu_{s}$. It can be observed that increasing the $r_{eff}$ and decreasing the a.r will not increase $\mu_{a}$. The geometrical cross section in this case will not counterbalance sufficient lower $Q_{abs}$. Particles with a.r$=3$ and $r_{eff}=$35 nm may be used for example only in scattering based detection systems.

Thus, from the optical imaging or photothermal perspective, larger AuNRs are
more favorable than the commonly used smaller AuNRs.

The length and width of this AuNR assuming a hemispherically capped cylinder geometry is 100 nm and 28.5 nm respectively. We qualify these particles as ``nanobig rods" for a clear identification further in discussion. We also name AuNRs with $r_{eff}$ smaller than 17.5 nm as being ``nanosmall rods" for differentiation in the discussion. As a thumb rule we define the ``nanobig rods"  as being those AuNRs with $r_{eff}$ greater than half of the mean free path of electrons in gold.

\begin{table}[!h]
\caption {Actual dimensions of gold nanorods occupying the classes
``nanosmall rods" and ``nanobig rods" with aspect ratios and effective volumes
that provide plasmon peaks at 800 nm.} \label{tab:dim}
\begin{center}
\begin{tabular}{c|c|c|c|c}
  \hline
$r_{eff}$& a.r & length (nm)&width(nm)&size class\\
\hline
11&4.0&50&12.5&``nanosmall rods"\\
&&&&\\
24&3.5&100&28.5&``nanobig rods"\\
\hline
\end{tabular}
\end{center}
\end{table}

We will further compare the commonly used ``nanosmall rods"
with the proposed ``nanobig rods". The dimensions of these particles are summarized in  Table~\ref{tab:dim}.

\subsection*{Circulation in the blood, extravasation and interstitial penetration}
\subsubsection{Circulation of AuNR in blood}

Upon intravenous administration of AuNRs, the particles are
distributed \emph{via} the vascular system to various organs and
tissues in the body. During this transport, the AuNRs interact
with various blood components such as cells and proteins.
Nanoparticles coated with hydrophilic, neutral polymers such as
PEG, possess relative stealth properties in the
blood~\cite{Niidome2006}. The steric stabilization of the
particles with the polymers prevents or minimizes adsorption of
proteins such as opsonin ligands, which could otherwise interact
with macrophage cell receptors and thereby mark a particle for
uptake. We have recently shown that PEGylation of AuNR prolongs
the blood circulation half-life of the particles to 19 hours,
while the non-PEG AuNR are trapped rapidly within minutes in the
liver and spleen~\cite{Maltzahn2009,Lankveld2011}.

Particles can be bioconjugated with specific antibodies to attach to targeted cells. However, it has been shown that the presence of
antibodies on the surface of particles increases their chances
of being recognized and engulfed by macrophages in the
RES (Reticulo Endothelial System)\cite{Koning2003,Koning2001a}. It was also shown that the uptake level of bioconjugated
nanoparticles  by macrophages is  proportional with loading of
antibodies on the surface ~\cite{Albanese2012}. Only at low densities of antibody coverage, sufficient target binding to tumor cells  can be achieved while maintaining minimal macrophage interaction.

 At this point, there is no literature comparing RES uptake of AuNR of various sizes; most studies have used the ``nanosmall rods" particles (65x11nm~\cite{Niidome2006}, 50 x15 nm \cite{Kogan2008}, 56x13 nm\cite{Wang2010}).

 Another issue with smaller nanoparticles represents their capacity to cross the Brain Blood Barrier (BBB). This phenomen is not desired if not specifically targeted. As discussed in recent publications \cite{Khlebtsov2011,Khlebtsov2011a, Terentyuk2009, Zhao2011} particles smaller than 20 nm can cross the BBB and can be found also in retinal layers. Larger particles such as ``nanobig rods" (width $>20$ nm) will have lower probability to cross the BBB.

 In the spleen, studies have shown that particles between 100 and 150 nm are more
efficiently filtered by the splenic bed~\cite{Moghimi1991}. At
this moment there is insufficient data to present clear evidence
for the preference of ``nanosmall rods" or ``nanobig rods" particles in this context.

\subsubsection{Extravasation}
Unlike normal vasculature, angiogenic blood vessels associated
with carcinoma, are in general poorly organized with chaotic
branching and are dilated, tortuous and leaky. The vessels can
have gaps as large as 10 $\mu$m between endothelial
cells~\cite{RKJain_ContRel1998, Seynhaeve2007, Koning2007}. Moreover, fenestration sizes vary from patient to patient, dependence on the type of cancer and change over time.

Solid tumors have interstitial hypertension which reduces the
convection of the particles across the vessel walls. The
progressively higher pressures from the periphery to the center of
of the tumor are due to the proliferation of cancer cells and
impaired lymphatics. For the NRs to infiltrate the tumor, the
particles will have to traverse this fluid interspersed
cross-linked network of collagen and elastin
fibers~\cite{RKJain_ContRel1998,RKJain_NatMed1998}. Movement in
this region is by diffusion and convection, however convective
transport progressively reduces towards the center of the tumor
due to the elevated pressures.

Experimental studies using 90 nm liposomes have shown that these
could penetrate only distances of 10-20 $\mu$m from
the microvasculature in mice, forming relatively static
perivascular clusters~\cite{Yuan1994}. It has been estimated that
such large particles could take months to traverse 1 mm of tumor
tissue, while smaller structures such as an IgG molecule with a
hydrodynamic radius of 5 nm would take a considerably lower 2-3
days~\cite{Kim2007103}. Thus, it is evident that smaller sizes of
the NRs will be favorable for interstitial penetration. Data
presented in literature on extravasation of nanoparticles shows that
larger particles tend to remain localized at the vascular bed and
do not extravasate deep into tumor\cite{Albanese2012}. This could make the case against ``nanobig rods" , since the spatial heterogeneity of particle distribution in tumors can affect the visualization of tumors by photoacoustic
methods by not revealing the entire tumoral volume. However, their presence at the tumor periphery can provide indications about the tumor size.

\subsubsection{Uptake by Target Cells}
When mAB-AuNRs  interact with the target cells, they can be taken
up by the cells \emph{via} receptor-mediated endocytosis (RME)~\cite{Chithrani2007,Chithrani.NanoLett2006}.
This process is initialized by clathrin, a cytoplasmic protein
which assists the formation of a vesicle around the particles,
through invagination of the plasma membrane. The vesicle, or early
endosome, is transported further to fuse with a lysosome where the
constituents are degraded and ultimately excreted. This process
can be fast and it usually takes less than an hour for a particle
to cross the membrane and to arrive in the
lysosome~\cite{Carmeliet2000}. A high surface area of the particle
can help the internalization process, firstly by increasing the
amount of antibodies that can be loaded on their surface, and
secondly by increasing the area projected on the surface of the
cell and thus the adhesion force as discussed in \cite{Best2012,Decuzzi2004,Decuzzi2008,Decuzzi2009}. The
internalization process is desirable in some therapeutic
applications because the drug molecules attached on their surface
can be released in cytoplasmic space \cite{Bardhan2011}. Another reason is
that the temperature fields generated by the particle during laser
irradiation can be confined in intracellular spaces localizing the
affected area close to the nucleus or other important cellular
organelles~\cite{Khlebtsov2006}. However, experiments in microsurgery using pulsed lasers have shown that
the damage to cells produced by explosion, cavitation or bubble formation is limited to the targeted cells~\cite{Huff2007a,Lukianova-Hleb2010,Chen2010d}.
For larger tumors, where not all the cells contain AuNR this type of therapy will not be successful to eradicate the tumor.
However, drug delivery can be still achieved \emph{in vivo} by short pulsed lasers~\cite{Braun2009,Jelveh2011} by cleaving the bond between the drug molecule and gold nanoparticle.

Chitrani \textit{et al}~\cite{Chithrani2007} showed that shorter
aspect ratios AuNRs with higher effective radii are internalized
in greater numbers than longer aspect ratio NRs, which is
supported by studies by Decuzzi and Ferrari~\cite{Decuzzi2008}. In
exocytosis (the excretion of foreign bodies by the cells), the
trend is reversed as the fraction of exocytosed particles is
higher for higher aspect ratio AuNRs. Consequently, shorter aspect
ratio NRs can stay longer in cells. Moreover, the optimum radius
of spherical nanoparticles for internalization was found to be
around 25 nm, both from experiment~\cite{Chithrani2007,Jiang2008}
and theory~\cite{Zhang2008, Gao2005}. New reports~\cite{Albanese2012} show that this optimum size actually is larger than 25 nm.  This suggests that the
``nanobig rods"  with an effective radius around 24 nm and a shorter
aspect ratio of 3.5 possess a higher propensity for cellular
internalization than the conventional ``nanosmall rods".

However, in the case of AuNR the internalization may not be desired because particles once engulfed
tend to form large aggregated structures~\cite{Ungureanu2011}. This situation will bring particles into plasmonic interaction range causing LP peaks to shift and broader depending on structure of the cluster.
\subsection*{Thermal response during pulsed laser irradiation}
Pulsed lasers can induce larger temperatures in nanoparticles
because heat will accumulate faster than losses across the
surfaces. The temperature rise of the particles can be
sufficient to cause vaporization of surrounding water/tissue
layers~\cite{Gonzalez2010, Lukianova-Hleb2010, Liu2009}. When
particle concentrations are high enough, bubble formation and
subsequent collapse can cause damage to cells. With high laser
intensities, the temperature rise can be so high that reshaping of
the NRs into shorter NRs and into spheres may occur. The melting
and/or fragmentation~\cite{Akchurin2008, Didychuk2009} causes a
drastic change in the optical properties, with the disappearance
of the LP peak in an ensemble collection of irradiated AuNRs.

The stability of the NRs in laser field depends on the melting
point of the particles, which in turns depends on their size and
coating~\cite{Link2000}. The energy (\emph{Q}$_{m}$) and temperature
required (\emph{T}$_{m}$) for melting of AuNRs can be calculated
using~\cite{Link2000,Qi2004}:

\begin{equation}\label{mel}
Q_{m}=\rho V(c_{p}[T_{NR}-T_{0})+ \Delta H_{f}]
  \label{eq:melt1}
\end{equation}

\begin{equation}\label{mel}
    T_{NR}=T_{b}(1-6\frac{r S}{8\pi r_{eff}^{3}})
     \label{eq:melt}
\end{equation}
where $\rho$ is bulk density (19300 kg m$^{-3}$), $c_{p}$ is heat
capacity (129 J kg$^{-1}$ K$^{-1}$), $\Delta H_{f}$ is enthalpy of
fusion(6.5 $\times$ 10$^{4}$ J kg$^{-1}$), $T_{b}$ is the bulk
melting temperature (1330 K), $T_{0}$ (310K) is the initial
temperature, $T_{NR}$ is the nanoparticle melting temperature and
$\emph{r}$ is the atomic radius (135 pm) of gold. \emph{V} is the
volume and $S$ the surface area of the AuNR before melting.
Equation \ref{eq:melt} accounts for the shape factor and for the
cohesive energy of the metallic particle~\cite{Qi2004}.

Using these equations, we obtain melting temperatures for
``nanosmall rods"  as 1281 K, and for ``nanobig rods"  as 1307 K. As
expected, the smaller particle possesses a lower temperature for
melting. The calculated energy (equation~\ref{eq:melt1}) required
to completely reshape ``nanobig rods" and ``nanosmall rods" into spheres
is 215 fJ and 20 fJ, respectively. Thus ``nanobig rods" particles are
more resistant to reshaping and can support higher laser powers,
making them better suited for use as contrast agents in
photoacoustic imaging.

\subsection*{Availability of synthetic methods}
The final criterion for the choice of nanoparticle dimensions is
whether the particles can be synthesized. The most well
established protocol, in terms of yield, monodispersity and fine
control of aspect ratio is the silver assisted seed-mediated
method of Nikoobakht and
El-Sayed~\cite{Nikoo.ChemMat2003,Khlebtsov.AO2005} that requires
the addition of gold seed to a growth solution with shape/size
directing surfactant CTAB and silver nitrate. Using different
AgNO$_{3}$ concentrations in the growth solution allows excellent
tuning of the a.r of the NRs, but only to a maximum of around 4. In
general these AuNRs fall under the ``nanobig rods" class (see
Table~\ref{tab:te}).

The method to grow AuNR using wet chemistry methods had actually
been pioneered by the Murphy group~\cite{Murphy2001}, but their
method did not use silver. Gold seed are made to initiate
nanocrystal growth in growth solutions with CTAB with slow
controlled reduction. The groups of Liz-Marzan and
Mulvaney~\cite{PerezJuste2004} showed that temperature, CTAB
concentrations, amount and sizes of the gold seed and decrease in
the reaction rate allows good control over the sizes and a.r of the
particles. Further
studies~\cite{Xia.AngewandteChemie2009,Marzan.CSR2008} also showed
that the a.r could be made to increase monotonously in size throughout the
growth process, which is not observed when silver is present
during reduction. Further the sizes of the particles are larger
than those which use the silver -assisted aproach and can be
described as ``nanobig rods" NRs (see Table~\ref{tab:te}).

\begin{table}[!h]
\caption {Typical sizes of AuNR synthesizable with different
protocols} \label{tab:te}
\begin{center}
\begin{tabular}{c|c|c|c|c|c|c}
  \hline
  reference & length (nm) & width(nm) & effective  & aspect  & position & class \\
  & & & radius (nm)& ratio & of LP (nm) &\\
  \hline
 \cite{Brioude2005}&  10-30&5-10&                            4-8 & 2.2-4 & 670-790 &
 ``nanosmall rods"
 \\
&&&&&&\\
\cite{Rayavarapu} &41-52&14-20&              11-15 & 2.3-3.6 &
675-850 & ``nanosmall rods" \\ &&&&&&\\
 \cite{PerezJuste2004} & 52-187&20-30    &                    15-34 & 2.2-6.7 & 724-1080 & ``nanobig rods"\\
  \hline
\end{tabular}
\end{center}
\end{table}

It should be admitted that methods which do not use silver result in large amounts of spheres as byproducts. However,
efficient separation of AuNR from mixtures of nanorods and
nanospheres can be performed using centrifugation utilizing
shape-dependent sedimentation behavior~\cite{Sharma2009}.

\section*{Concluding remarks}

 We have identified
a size class, ``nanobig rods", which has several advantages compared to
their smaller counterparts, ``nanosmall rods", which are used
conventionally. The discussed performance of these two size
classes for various features is summarized in Table 3.

\begin{table}[!h]
\caption {Performance of the two NR classes in the selection
criteria \label{tab:tem}}
\begin{center}
\begin{tabular}{l|c|c}
  \hline

Feature &  ``nanobig rods" & ``nanosmall rods" \\
  \hline
LP at 800 nm &  + & + \\
$C_{abs}$ at LP peak &  ++ & + \\
$C_{sca}$ at LP peak &  ++ & + \\
Circulation in blood &  ? & ? \\
extravasation at tumor &  + & ++ \\
interstitium transport &  + & ++ \\
target cell uptake &  ? & ? \\
lower number for contrast effect &  ++ & + \\
thermal stability &  ++ & + \\
quality of synthesis &  + & ++ \\
  \hline
\end{tabular}
\end{center}
\end{table}

Some performance indicators can be conflicting and a judicious
balance between the different criteria has to be found. The extent
of mAb loading on a particle, which depends on the surface area,
e.g. will improve the adhesion to the target cells. On the other
hand, enhanced immunogenicity of the particle with more mAbs may
increase uptake by the RES and thus lower the final dose arriving in the
tumor. Moreover, as shown in\cite{Kirpotin2006}, the bioconjugation improve the internalization rather than extravasation.

 Finally,
modulation of the microenvironment of the tumors may improve the
uptake of the AuNRs. Some examples of vascular manipulation have been described to improve delivery of drugs and/or drug containing nano-carriers to solid tumors. Transient normalization~\cite{K.Jain2005} of the abnormal structure of
tumor vasculature, is known to improve perfusion and thereby drug delivery. The use of vaso-dilatation factors~\cite{Sonveaux2008}, or normalization approaches that increase vascular permeability for instance by using growth factors or cytokines~\cite{Seynhaeve2007}, or heat~\cite{Kong2000,Li2010} are known to increase accumulation and transport of NPs. A final approach of vascular manipulation is the approach that aims at lowering the interstitial hypertension using lytic enzymes\cite{Jain1990,RKJain_ContRel1998}.

Additional experiments and theoretical calculations are necessary to show which among ``nanobig rods" and ``nanosmall rods" particles are better suited for biological applications. Techniques similar to Particle Swarm Optimization (PSO)~\cite{Kessentini2011} may be used to optimize the size of gold nanorods for biological applications.  However, if we take into account only the optical and thermal properties, the ``nanobig rods" particles have better properties for biological applications than ``nanosmall rods" ones.

\section{Acknowledgements}
This work is funded through the thrust area program NIMTIK of the
University of Twente; through the PRESMITT project (IPD067771) of
the SenterNovem program IOP Photonic Devices; and by the
Nederlandse Wetenschappelijk Organisatie (NWO) and Stichting
Technische Wetenschappen (STW) through project TTF 6527.

\section*{References}
\bibliography{nanobig1811}
\bibliographystyle{nano3}
%

\end{document}